\documentclass[12pt]{article}
\usepackage{graphicx}

\newcommand\pubnumber{SNSN-323-63}
\newcommand\pubdate{\today}

\def\CERN{CERN\\
Geneva, Switzerland}

\def\Title#1{\begin{center} {\Large #1 } \end{center}}
\def\Author#1{\begin{center}{ \sc #1} \end{center}}
\def\Address#1{\begin{center}{ \it #1} \end{center}}

\newcommand\pubblock{\rightline{\begin{tabular}{l} \pubnumber\\
         \pubdate  \end{tabular}}}
\newenvironment{Abstract}{\begin{quotation}  }{\end{quotation}}
\newenvironment{Presented}{\begin{quotation} \begin{center} 
             PRESENTED AT\end{center}\bigskip 
      \begin{center}\begin{large}}{\end{large}\end{center} \end{quotation}}
\def\Acknowledgements{\bigskip  \bigskip \begin{center} \begin{large}
             \bf ACKNOWLEDGEMENTS \end{large}\end{center}}
\def\beq{\begin{equation}}
\def\eeq#1{\label{#1}\end{equation}}
\def\eeqn{\end{equation}}

\def\beqa{\begin{eqnarray}}
\def\eeqa#1{\label{#1}\end{eqnarray}}
\def\eeqan{\end{eqnarray}}

\let\bar=\overbar

\def\Dslash{\not{\hbox{\kern-4pt $D$}}}
\def\dslash{\not{\hbox{\kern-2pt $\del$}}}

\def\msb{{\bar{\ssstyle M \kern -1pt S}}}

\begin{document}
\begin{titlepage}
\pubblock

\vfill
\Title{Measurement of $CP$ observables in $B_s^0 \rightarrow D_s^{\mp}K^{\pm}$ at LHCb}
\vfill
\Author{Vladimir Gligorov\\ On behalf of the LHCb collaboration}
\Address{\CERN}
\vfill
\begin{Abstract}
The time-dependent $CP$-violating observables accessible through $B_s^0 \rightarrow D_s^{\mp}K^{\pm}$ decays have been measured for the first time
using data corresponding to an integrated luminosity of 1 $fb^{-1}$ collected in 2011 by the LHCb detector.
The $CP$-violating observables are found to be: 
$C_f = 0.53 \pm 0.25 \pm 0.04$,
$A^{\Delta\Gamma}_{f} = 0.37 \pm 0.42 \pm 0.20$, 
$A^{\Delta\Gamma}_{\overline{f}} = 0.20 \pm 0.41 \pm 0.20$, 
$S_f = −1.09 \pm 0.33 \pm 0.08$,
$S_{\overline{f}} = −0.36 \pm 0.34 \pm 0.08$,
where the first uncertainty is statistical and the second systematic. 
Using these observables, the CKM angle $\gamma$ is determined to be $(115_{-43}^{+28})^\circ$ modulo $180^\circ$ at $68\%$ CL,
where the uncertainty contains both statistical and systematic components.
\end{Abstract}
\vfill
\begin{Presented}
The 8th International Workshop on the CKM Unitarity Triangle (CKM 2014)\\
Vienna, Austria, September 8-12, 2014
\end{Presented}
\vfill
\end{titlepage}
\def\thefootnote{\fnsymbol{footnote}}
\setcounter{footnote}{0}

\section{Introduction}
Matter-antimatter asymmetry ($CP$ violation) in weak interactions is described by a single,
irreducible phase in the Cabibbo-Kobayashi-Maskawa (CKM) quark mixing matrix \cite{CKM1,CKM2}. 
As this is a $3\times 3$ unitary, hermitian, matrix, it can be represented as a ``Unitarity Triangle'' in the complex plane.
Since the matter-antimatter asymmetry in the Standard Model is too small~\cite{Huet:1994jb} to account
for the disappearance of antimatter following the Big Bang, it is reasonable to suppose that the
Standard Model picture of $CP$ violation is not self-consistent and breaks down at some level.
By experimentally overconstraining the Unitarity Triangle, we are therefore directly
probing the energy scale of potential physics beyond the Standard Model.

The time-dependent decay rates of the
$|B_s^0(t=0)\rangle$ and $|\overline{B}_s^0(t=0)\rangle$ flavour eigenstates to final state $f$ are:
\begin{center}
 $\frac{d\Gamma_{B_s^0 \to f}(t)}{dt} \propto e^{-\Gamma_st}[cosh(\frac{\Delta \Gamma_s t}{2}) 
 +A_{f}^{\Delta \Gamma}sinh(\frac{\Delta \Gamma_s t}{2}) +C_f cos(\Delta m_s t)-S_f sin(\Delta m_s t)]$, \\ 
\vspace{0.25cm}
  $\frac{d\Gamma_{\overline{B}_s^0 \to f}(t)}{dt} \propto e^{-\Gamma_st}[cosh(\frac{\Delta \Gamma_s t}{2}) 
 +A_{f}^{\Delta \Gamma}sinh(\frac{\Delta \Gamma_s t}{2}) -C_f cos(\Delta m_s t)+S_f sin(\Delta m_s t)]$. 
\end{center}

Similar decay rates hold for the conjugate processes.
In the case where $f \equiv D_s^{-}K^{+}$, the four decay rates give five independently measureable $CP$-violating observables (``$CP$ observables'' henceforth), which are
related to $r_{D_sK} \equiv  |A(\overline{B_s}^0 \to D_s^- K^+)/A(B_s^0 \to D_s^- K^+)|$,
the ratio of the magnitudes of the interfering diagrams, as well as
the strong phase difference $\delta$ and the weak phase difference $\gamma -2\beta_s$:

\begin{center}
\label{eq:observables}
    $C_{f}   = \frac{1-r_{D_sK}^2}{1+r_{D_sK}^2}$, 
    $A_{f}^{\Delta \Gamma}   = \frac{-2 r_{D_sK} \cos(\delta-(\gamma-2\beta_s))}{1+r_{D_sK}^2}$,
    $A_{\overline{f}}^{\Delta \Gamma}  = \frac{-2 r_{D_sK} \cos(\delta+(\gamma-2\beta_s))}{1+r_{D_sK}^2}$,\\ 
\vspace{0.25cm}
    $S_{f}   = \frac{2 r_{D_sK}\sin(\delta-(\gamma-2\beta_s))}{1+r_{D_sK}^2}$,
    $S_{\overline{f}}  = \frac{-2 r_{D_sK}\sin(\delta+(\gamma-2\beta_s))}{1+r_{D_sK}^2}$,
\end{center}

where $\beta_s \equiv \arg(-V_{ts}V_{tb}^{*}/V_{cs}V_{cb}^{*})$. These observables can therefore
be used to measure $\gamma$, an angle of the Unitarity Triangle, with negligible~\cite{Brod:2013sga} 
theoretical uncertainty.

\section{Cancellation of ambiguities}

\begin{figure}[ht]
\centering
\includegraphics[height=2.5in]{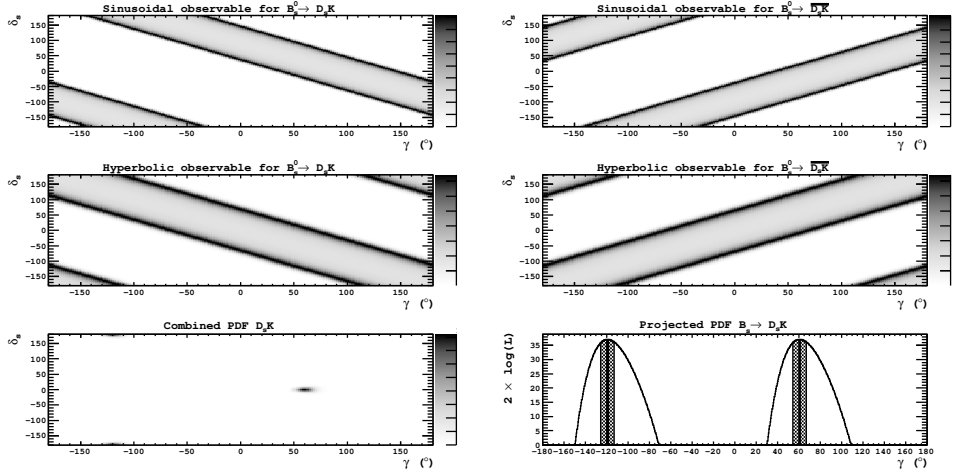}
\caption{Reproduced from~\cite{gligckm2010}. The top four plots show the likelihoods of the $CP$ observables with 31k signal events and
LHCb MC performance~\cite{Akiba:2008zza}. The bottom left plot
shows the combined likelihood in the $\gamma-\delta_s$ plane and the bottom right the projection onto $\gamma$, where the hatched area is the 1$\sigma$
region and the dark vertical line the central value. All but two of the ambiguities are excluded.}
\label{fig:gammaFromDsK_10fb}
\end{figure}

As discussed in~\cite{Fleisher}, the fact that $\Delta\Gamma_s$ is relatively large makes both the sinusoidal and hyperbolic $CP$ 
observables in $B_s^0 \rightarrow D_s^{\mp}K^{\pm}$ measurable and hence results in only a twofold ambiguity on the measured value of
the CKM angle $\gamma$ and the strong phase difference $\delta$. In order to illustrate this point, it is useful to consider the constraints on $\gamma$ due to
each of the observables listed in Eq.~\ref{eq:observables}. These are illustrated in Fig.~\ref{fig:gammaFromDsK_10fb}, which clearly shows
how the diagonal staggering of the sinusoidal and hyperbolic constraints in the $\delta-\gamma$ plane cancels all but one of
the ambiguous solutions. 

\section{Event selection}

The analysis uses a datasample corresponding to an integrated luminosity of 1 $fb^{-1}$
collected by LHCb detector in 2011. The full description of detector can be found in ~\cite{detector}. 
The trigger~\cite{lhcbtrigger} consists of a hardware stage, based on information from the 
calorimeter and muon systems, followed by a software stage, in which all charged particles
with $p_T > 500$ MeV are reconstructed, and a multivariate algorithm~\cite{Gligorov:2012qt} is used to select
displaced vertices compatible with the decay of a $b$-hadron.

The $D_s^-$ particle is reconstructed in three decay modes: 
$D_s^-  \rightarrow K^-K^+\pi^-$, $D_s^-  \rightarrow K^-\pi^+\pi^-$, and $D_s^-  \rightarrow \pi^-\pi^+\pi^-$. 
These $D_s^-$ candidates are subsequently combined with a fourth particle, referred to as the
``companion'', to form $B_s^0 \rightarrow D_s^{\mp}K^{\pm}$ and $B_s^0 \rightarrow D_s^{-}\pi^{+}$ candidates.
The flavour-specific Cabibbo-favoured decay mode $B_s^0 \rightarrow D_s^{-}\pi^{+}$ is used as a control channel
for the analysis, in particular for optimizing the selection and constraining the decay-time-dependent selection efficiency.

The different $D_s^-$ final states are distinguished by a combination of particle identification information
from LHCb's Ring Imaging Cherenkov (RICH) subdetectors and kinematic vetoes.
This selection also strongly suppresses cross-feed and peaking backgrounds from other misidentified
decays of $b$-hadrons to $c$-hadrons. 

\section{Multivariate fit to $B_s^0 \rightarrow D_s^{\mp}K^{\pm}$ and $B_s^0 \rightarrow D_s^{-}\pi^{+}$}
\label{sec:mdfit}
The signal and background yields in the $B_s^0 \rightarrow D_s^{\mp}K^{\pm}$
and $B_s^0 \rightarrow D_s^{-}\pi^{+}$ channels are determined using a three-dimensional simultaneous extended maximum likelihood fit in
the $B_s^0$ mass, the $D_s^-$ mass, and the log-likelihood
difference $L(K/\pi)$ between the pion and kaon hypotheses for the companion particle. 
Correlations between the fitting variables are shown to be negligible using simulated events.

\begin{figure}
 \centering
 \includegraphics[width=.32\textwidth]{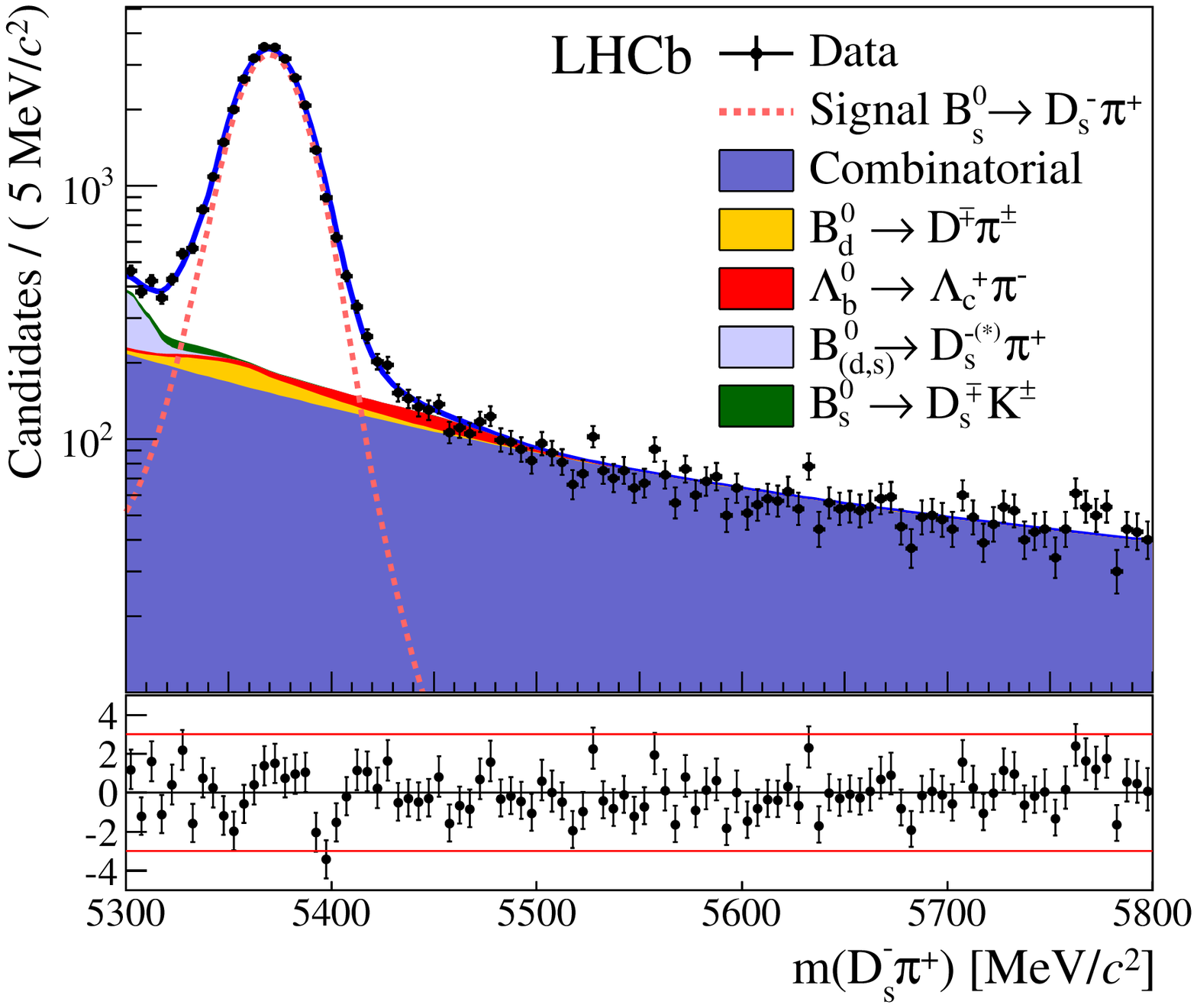}
 \includegraphics[width=.32\textwidth]{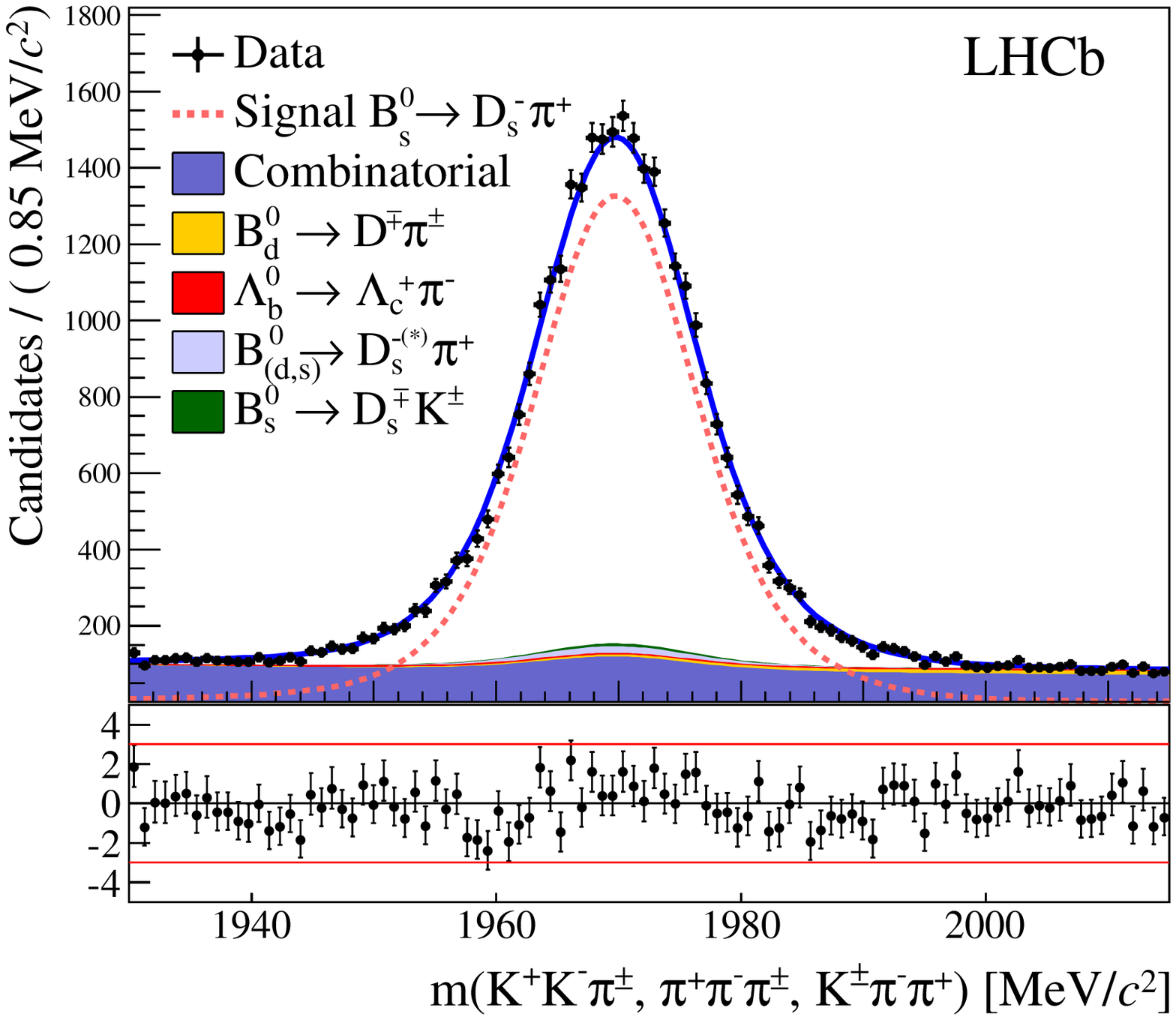}
 \includegraphics[width=.32\textwidth]{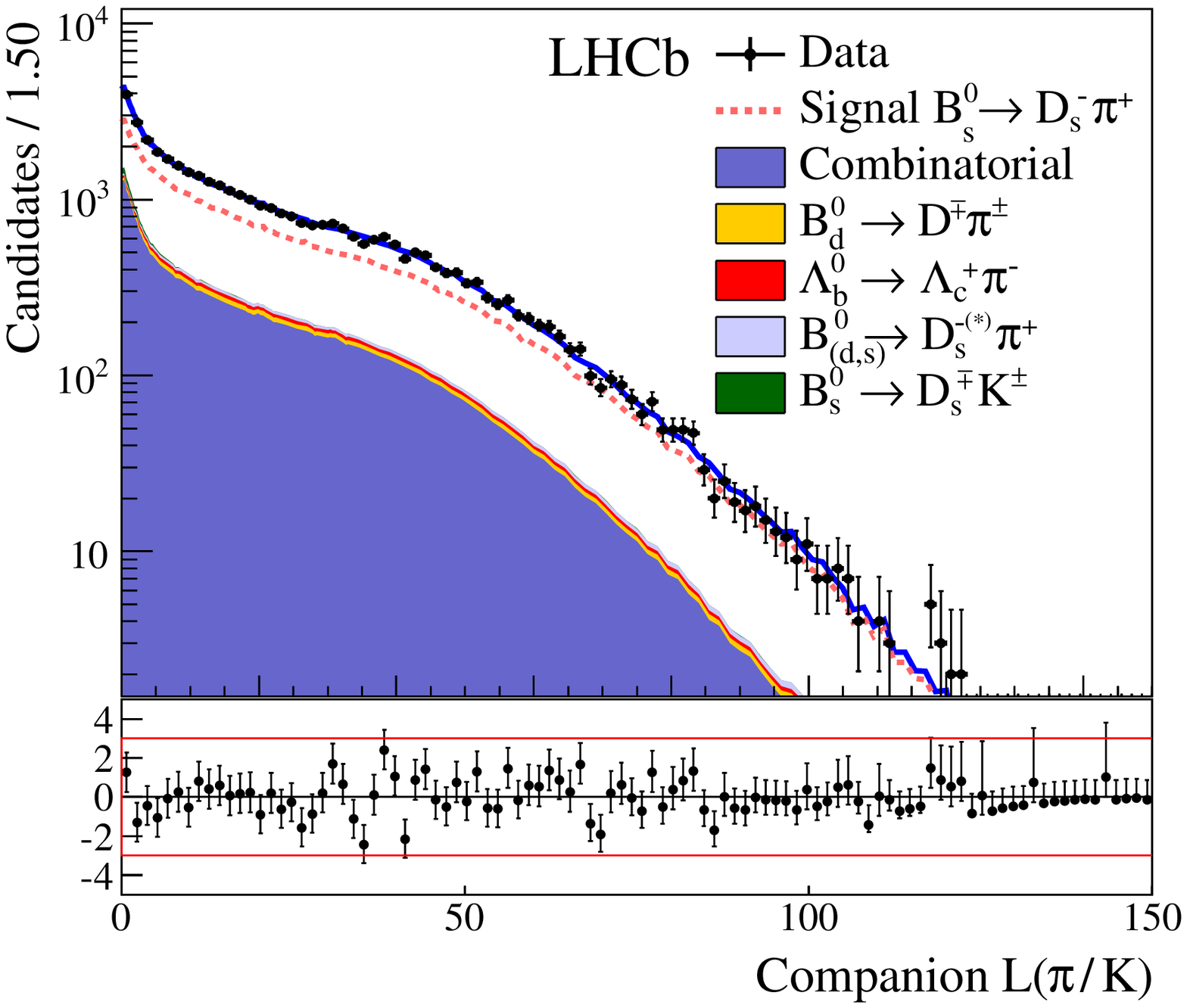}
 \caption{Multivariate fit to all $B_s^0 \rightarrow D_s^{-}\pi^{+}$ candidates.% for all $D_s^-$ decay modes combined. 
 Left to right: distributions of candidates in $B_s^0$ mass, $D_s^-$ mass, companion PID log-likelihood difference.
}
 \label{fig:mdfit_dspi}
\end{figure}

\begin{figure}
 \centering
 \includegraphics[width=.32\textwidth]{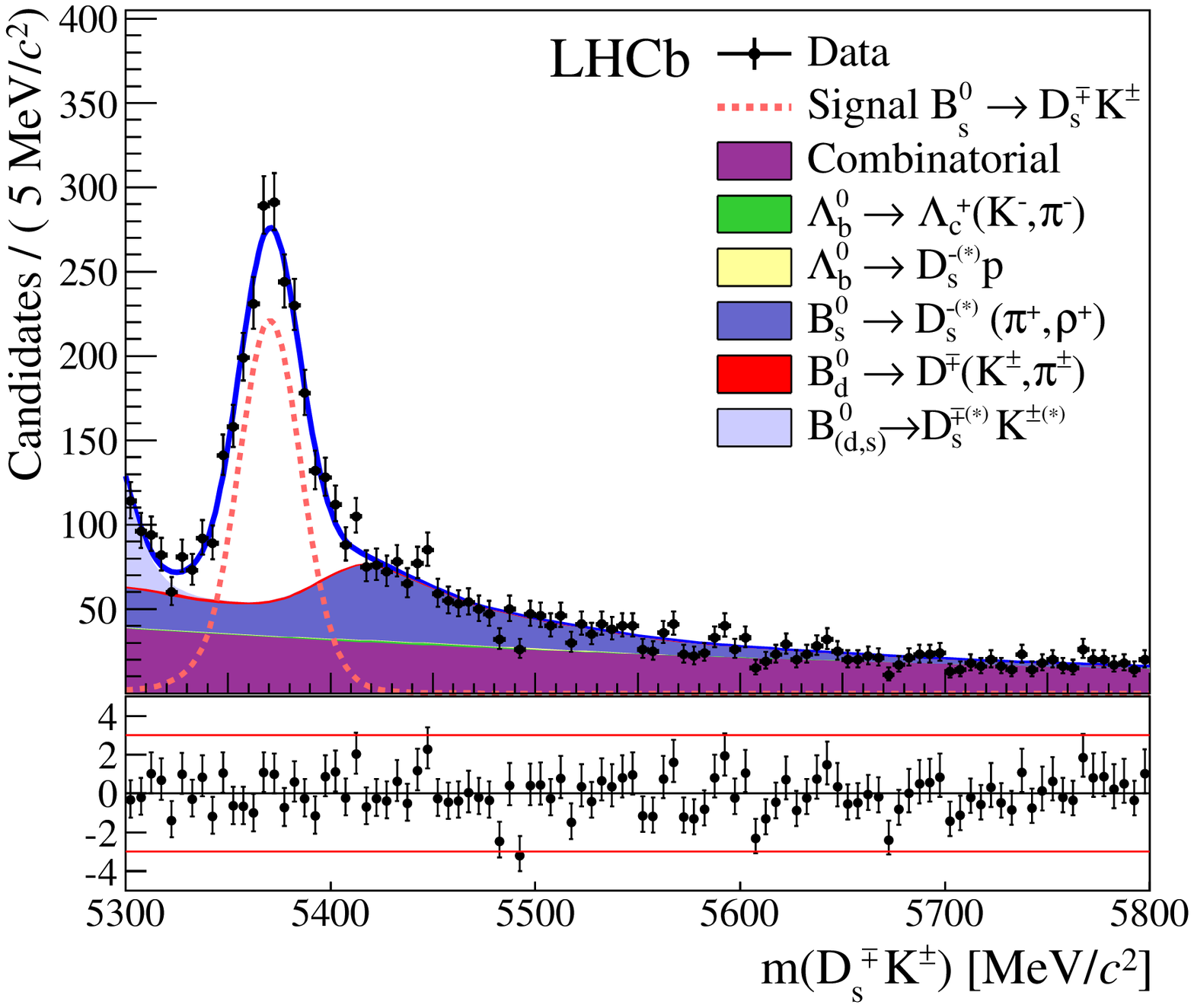}
 \includegraphics[width=.32\textwidth]{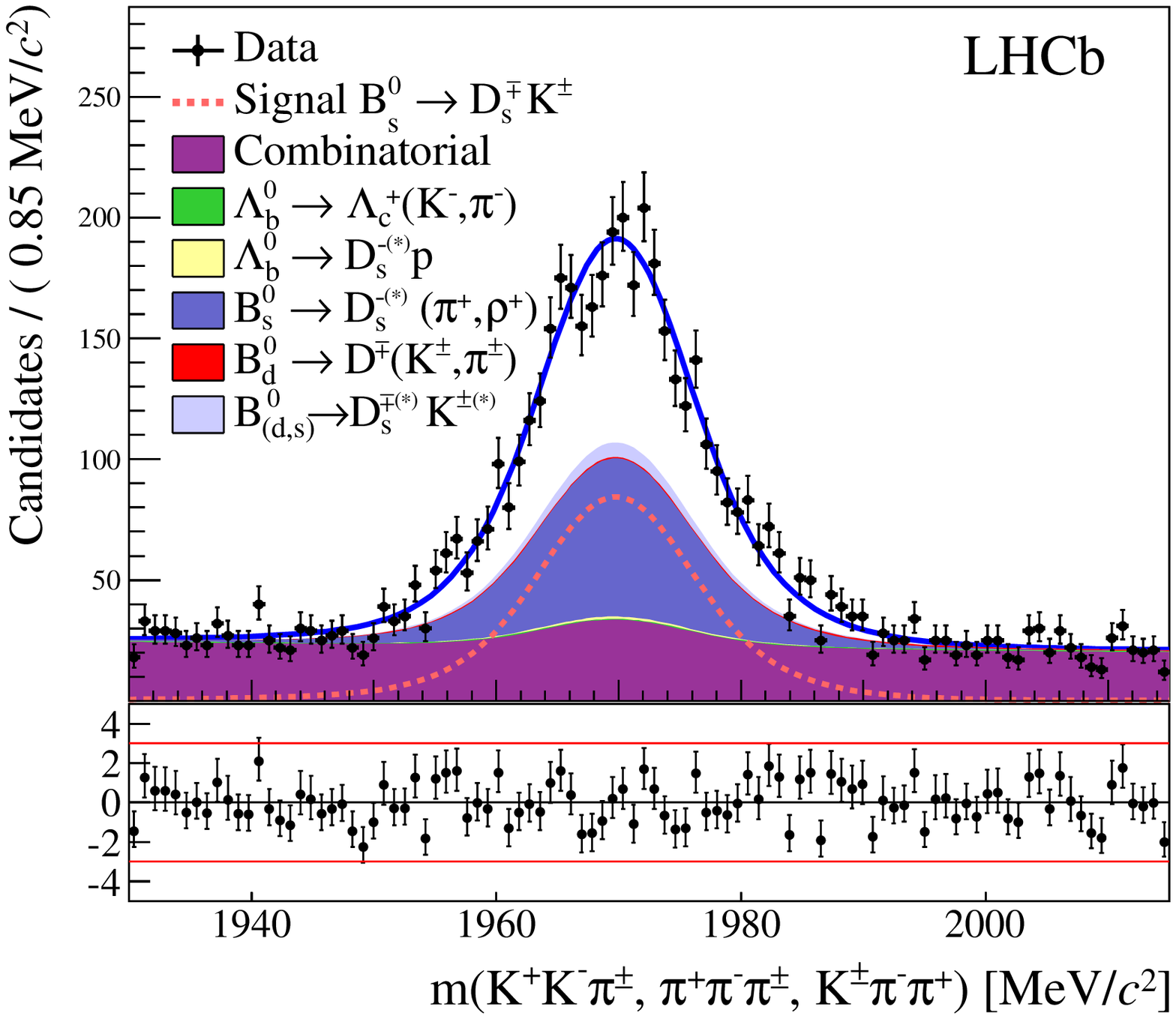}
 \includegraphics[width=.32\textwidth]{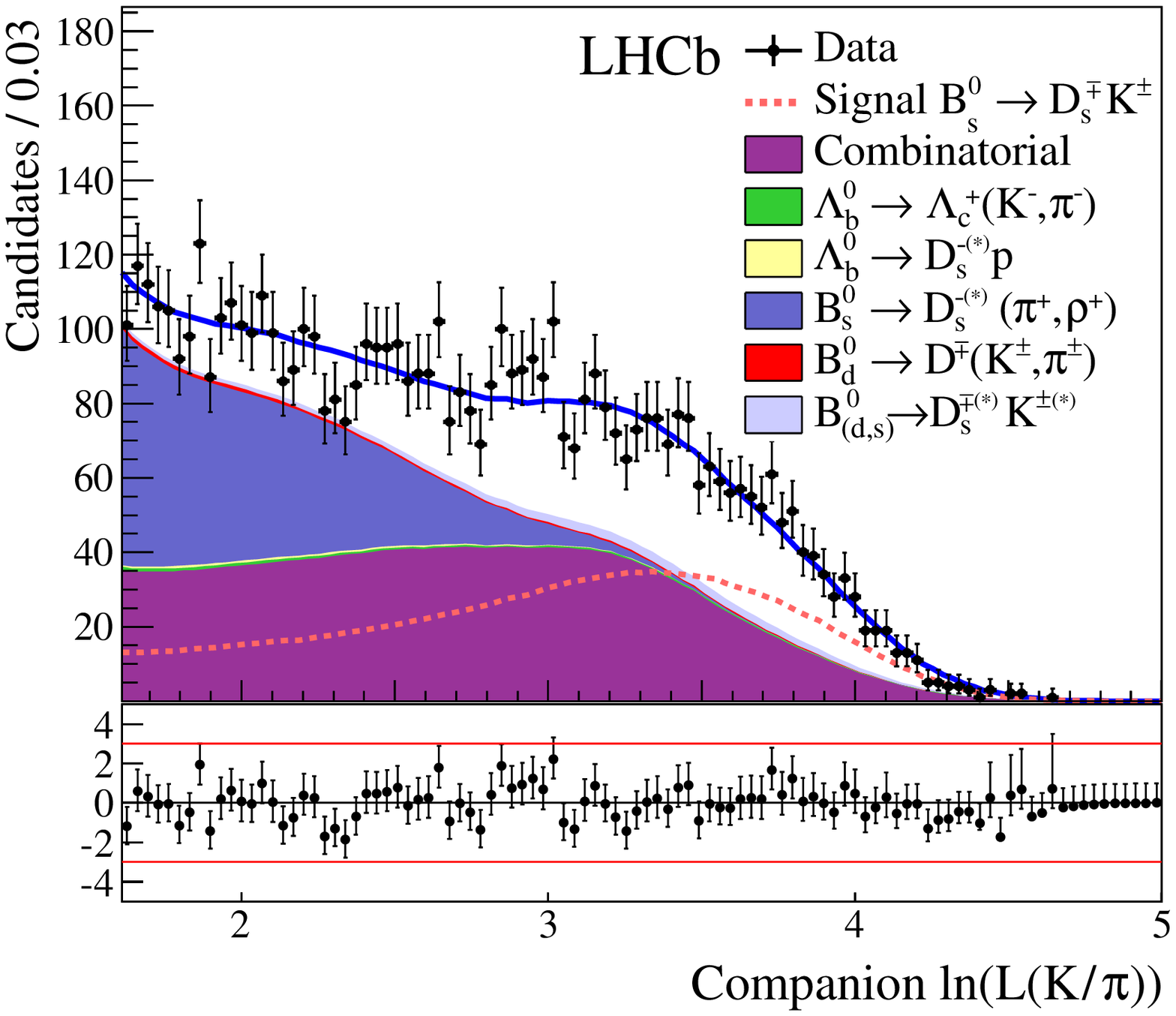}
 \caption{Multivariate fit to all $B_s^0 \rightarrow D_s^{\mp}K^{\pm}$ candidates.% for all $D_s^-$ decay modes combined. 
 Left to right: distributions of candidates in $B_s^0$ mass, $D_s^-$ mass, companion PID log-likelihood difference.}
 \label{fig:mdfit_dsk}
\end{figure}

The dominant backgrounds are random combinations of $D_s^{-}$ mesons with pions or kaons, partially reconstructed decays of the type $B_s^0 \rightarrow D_s^{-}(\pi,K)^{+} X$, and decays of $B^0$ and $\Lambda_b$ hadrons in which the $D^+$ or $\Lambda_c$ candidates are misidentified as $D_s^{-}$ candidates.
Most background yields float in the fit, with the exception of modes with yields  below $2\%$ of the signal yield 
which are fixed from known branching fractions and relative efficiencies measured using simulated events.
%Almost all background yields are left free to float, 
%however the backgrounds whose yields are below $2\%$ of the signal yield 
%are fixed from known branching fractions and relative efficiencies measured using simulated events.                                                                                                                                                                               
The multivariate fit results in a signal yield of $28\,260\pm 180$ $B_s^0 \rightarrow D_s^{-}\pi^{+}$ and $1770 \pm 50$ $B_s^0 \rightarrow D_s^{\mp}K^{\pm}$ decays,
shown in Fig.~\ref{fig:mdfit_dspi} and Fig.~\ref{fig:mdfit_dsk}, respectively.
%An effective purity of $85\%$ for $B_s^0 \rightarrow D_s^{-}\pi^{+}$ and $74\%$ for $B_s^0 \rightarrow D_s^{\mp}K^{\pm}$ are measured.
The multivariate fit is checked for biases using large
samples of data-like pseudoexperiments, and none are found.

\section{Inputs to the time-dependent fit}

The measurement of the sinusoidal components of the $B_s^0 \rightarrow D_s^{\mp}K^{\pm}$ decay rates requires the determination (``tagging'')
of the initial flavour of the $B_s^0$ meson. The performance of the LHCb flavour tagging algorithms is
described in detail in \cite{LHCbTagging}. This analysis uses two types of taggers: opposite side, which infer
the production flavour of the $B_s^0$ meson by partially reconstructing the other $b$-hadron produced in the $pp$ collision;
and same-side, which infer the production flavour of the $B_s^0$ meson by finding a charged kaon produced in the same
fragmentation chain. The total tagging efficiency is $67.53\%$ and the total effective tagging power is $5.07\%$.

The decay-time of the $B_s^0$ candidate is computed using a kinematic fit which constrains the mass of the $D_s^-$ meson to the world-average
value, as well as constraining the $B_s^0$ candidate to point to the associated $pp$ collision vertex. This fit also returns an estimated
per-event decay-time uncertainty, which is used as an observable when fitting to the decay rates in order to maximize sensitivity. The estimated
decay-time uncertainty is calibrated using prompt $D_s^-$ mesons which are combined with a random track and
kinematically weighted to give a sample of ``fake $B_s^0$'' candidates, and the scale factor is found to be $1.37\pm0.10$.

Because the hyperbolic $CP$ observables in $B_s^0 \rightarrow D_s^{\mp}K^{\pm}$ are fully correlated with the decay-time acceptance,
it must be independently measured and fixed in the fit. This is done using the known value of $\Gamma_s$ and
the $B_s^0 \rightarrow D_s^{-}\pi^{+}$ control channel. The obtained acceptance is corrected by the acceptance ratio of 
$B_s^0 \rightarrow D_s^{\mp}K^{\pm}$ and $B_s^0 \rightarrow D_s^{-}\pi^{+}$ found in simulation. In order to help fit stability and speed,
the decay-time acceptance is implemented using analytically integrable spline polynomials~\cite{Manuel}.
%In the above $B_s^0 \rightarrow D_s^{-}\pi^{+}$ fit $\Delta m_s$ is floating, giving value of $17.772 \pm 0.022$ $ps^{-1}$
%which is in excellent agreement with the published LHCb measurement
%of $\Delta m_s = 17.768 \pm 0.023\pm0.006$ $ps^{-1}$~\cite{deltaMS}. The result of $B_s^0 \rightarrow D_s^{-}\pi^{+}$ fit
%is shown in Fig.~\ref{fig:dspitime}.

Finally, the following paremeters are fixed from independent measurements~\cite{deltaMS,PDG,Gamma}:
$\Gamma_s = (0.661 \pm 0.007)ps^{-1}$,
$\Gamma_{\Lambda_b^0} = (0.676 \pm 0.006)ps^{-1}$,
$\Delta \Gamma_s = (0.106 \pm 0.013)ps^{-1}$,
$\Gamma_d = (0.658 \pm 0.003)ps^{-1}$,
$\Delta m_s = (17.768 \pm 0.024)ps^{-1}$,
$\rho(\Gamma_s,\Delta \Gamma_s) = -0.39$,\\
where $\rho(\Gamma_s,\Delta \Gamma_s)$ is the correlation between these two measurements, $\Gamma_{\Lambda_b^0}$ the $\Lambda_b^0$ decay-width,
$\Gamma_d$ the $B_d^0$ decay width, and $\Delta m_s$ the $B_s^0$ oscillation frequency. 

\section{Time-dependent fit to $B_s^0 \rightarrow D_s^{\mp}K^{\pm}$}

The determination of the $CP$ observables is performed using two different unbinned maximum likelihood fits.
In the first (\hspace{-0.125em}\textit{cFit}) all signal and background time distributions are
described. In the second (\hspace{-0.125em}\textit{sFit}) the background is statistically subtracted
using the \textit{sPlot} technique~\cite{sPlot} and only the signal time distributions
are described. The signal decay-time model is identical in the two fitters.

Decay-time PDFs for both signal and background components account for flavour tagging, are convolved with a single Gaussian
representing the per-candidate decay-time resolution, and are multiplied by the decay-time acceptance.
In the \textit{sFit} approach the signal $B_s^0 \rightarrow D_s^{\mp}K^{\pm}$ model is fitted to the three time-dependent observables: 
decay time, decay-time uncertainty, and predicted mistag.
In order to optimally discriminate against background,
the \textit{cFit} performs a six-dimensional fit to the time-dependent observables and the three observables
used in the multivariate fit. 

\begin{figure}[htb]
 \centering
 \includegraphics[width=.45\textwidth]{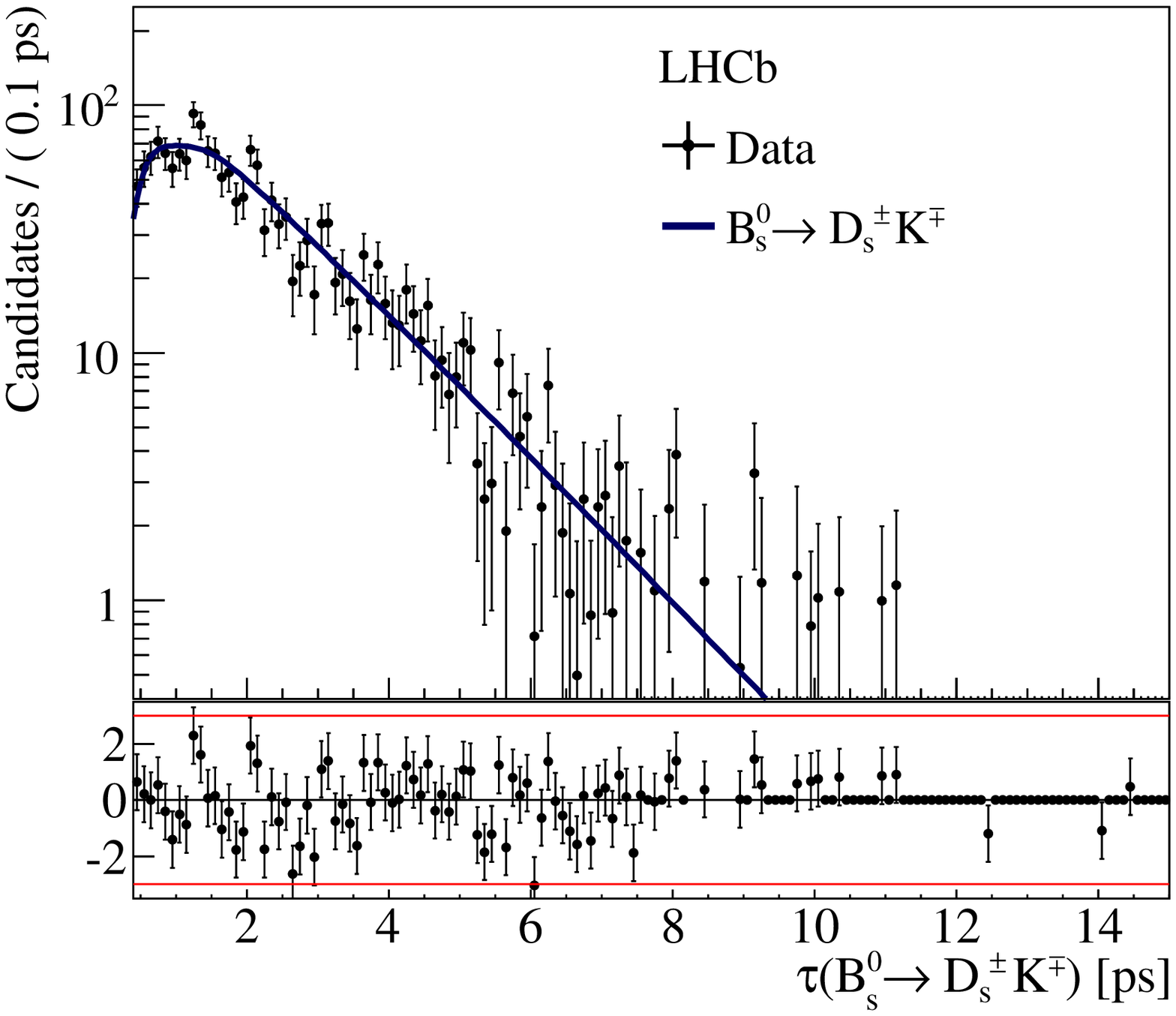}
 \includegraphics[width=.45\textwidth]{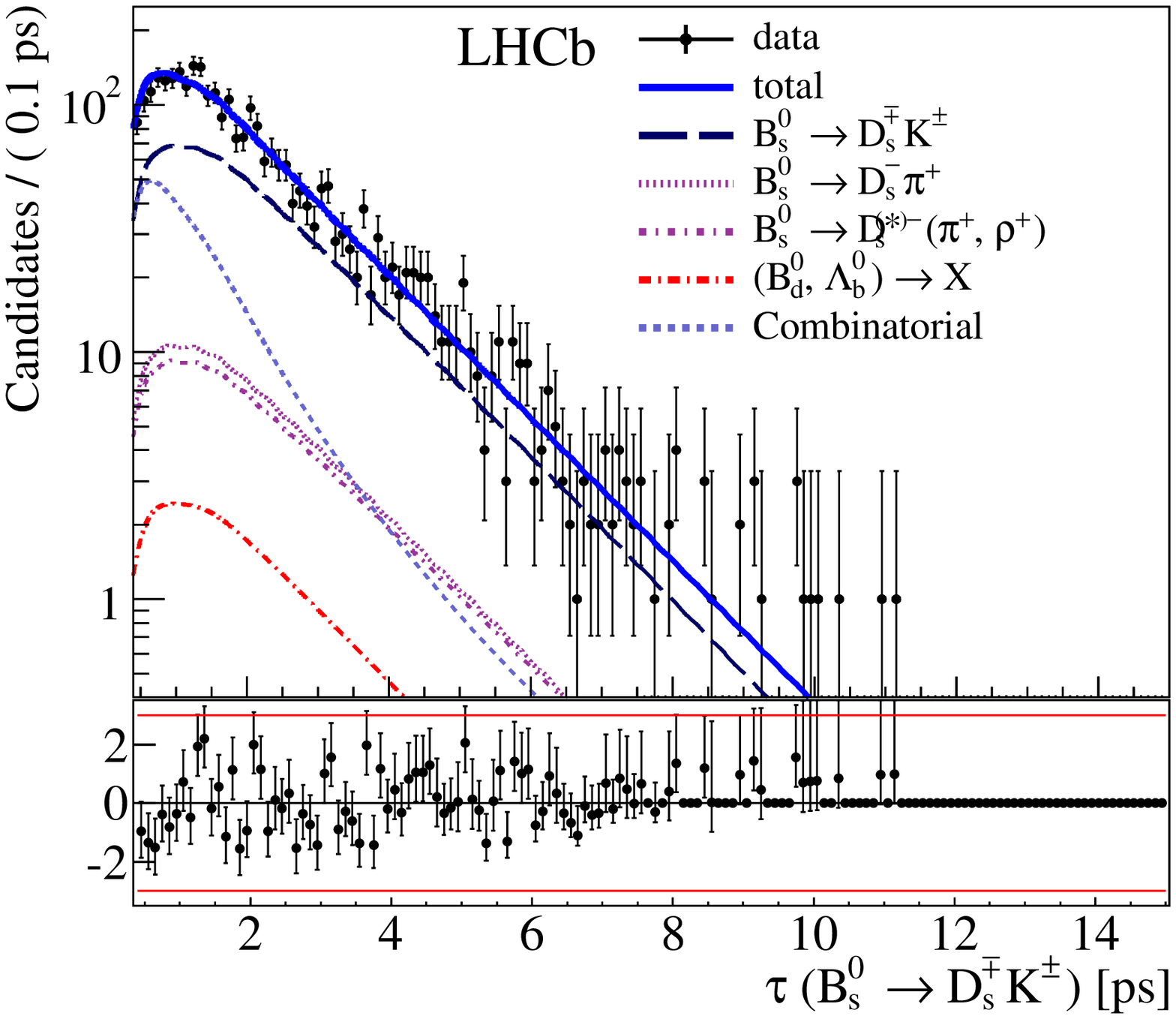}
 \caption{The (top) \textit{sFit} and (bottom) \textit{cFit} to the $B_s^0 \rightarrow D_s^{\mp}K^{\pm}$ candidate decay-time.}
 \label{fig:time}
\end{figure}

The results of the \textit{cFit} and \textit{sFit} for the $CP$ observables are given in Table~\ref{tab:time}, and shown in Fig.~\ref{fig:time}.
Systematic uncertainties divide into three kinds: uncertainties from the fixed parameters, 
uncertainties from the limited knowledge of the decay-time resolution and acceptance,
and uncertainties related to fit biases. The first two are estimated
using large sets of simulated pseudoexperiments, in which the relevant parameters are
varied within their uncertainty. The third is computed by splitting the data into independent subsamples, repeating the
entire analysis chain for each one, and comparing the weighted average of the results to the nominal result.
The measurement is statistically limited, and the largest systematic uncertainties are on the hyperbolic observables, due to the limited 
knowledge of $\Gamma_s$, $\Delta\Gamma_s$, and the decay-time acceptance.

\begin{table}[htbp]
\centering
\caption{Fitted $CP$ observables for (left) \textit{sFit} and (right) \textit{cFit}. The first uncertainty is statistical and the second is systematic.}
\vspace{0.25cm}
\label{tab:time}
  \begin{tabular}{c c c} 
      \hline 
  	Parameter &  sFit fitted value & cFit fitted value \\	         
  	\hline 
  	$C_{f}$                               &  $\phantom{-}0.52 \pm 0.25 \pm 0.04$  &  $\phantom{-}0.53 \pm 0.25 \pm 0.04$ \\
  	$S_{f}$                               &  $-0.90 \pm 0.31 \pm 0.06$            &  $-1.09 \pm 0.33 \pm 0.08$ \\
  	$S_{\overline{f}}$                    &  $-0.36 \pm 0.34 \pm 0.06$            &  $-0.36 \pm 0.34 \pm 0.08$  \\
  	$A_{f}^{\Delta \Gamma}$               &  $\phantom{-}0.29 \pm 0.42 \pm 0.17$  &  $\phantom{-}0.37 \pm 0.42 \pm 0.20$ \\
  	$A_{\overline{f}}^{\Delta \Gamma}$    &  $\phantom{-}0.14 \pm 0.41 \pm 0.18$  &  $\phantom{-}0.20 \pm 0.41 \pm 0.20$ \\
  	\hline 
  \end{tabular}
\end{table}

As the \textit{cFit} and \textit{sFit} sensitivities are very similar, the nominal result was randomly chosen to be the \textit{cFit}.

\section{Determination of the CKM $\gamma$ angle}
\label{sec:gamma}

The measurement of the $CP$ observables is interpreted in terms of $\gamma - 2\beta_s$ by maximising 
\begin{center}
$\mathcal L(\vec \alpha) = exp(-\frac{1}{2}(\vec A(\vec \alpha)- \vec A_{obs})^T V^{-1}(\vec A(\vec \alpha)- \vec A_{obs}))$, $\quad$ (5)
\end{center}
where $\overrightarrow{\alpha} = (\gamma,\phi_s,r_{D_sK},\delta)$ is the vector of the physics
parameters, $\vec A$ is the vector of observables, $\vec A_{obs}$ is the vector of the measured
$CP$ observables and $V$ is the experimental (statistical and systematic)
covariance matrix. 

The value of mixing phase is constrained by $-2\beta_s = \phi_s = 0.01 \pm 0.07(stat)\pm 0.01(syst)$ $rad$ 
from the LHCb measurement of $B_s^0 \to J/\phi K^+K^-$ and $B_s^0 \to J/\phi \pi^+\pi^-$ decays~\cite{Gamma}.
This assumes that penguin pollution and BSM contributions are negligible, which is certainly a good approximation at the
present statistical sensitivity.

\begin{figure}
 \centering
 \includegraphics[width=.30\textwidth]{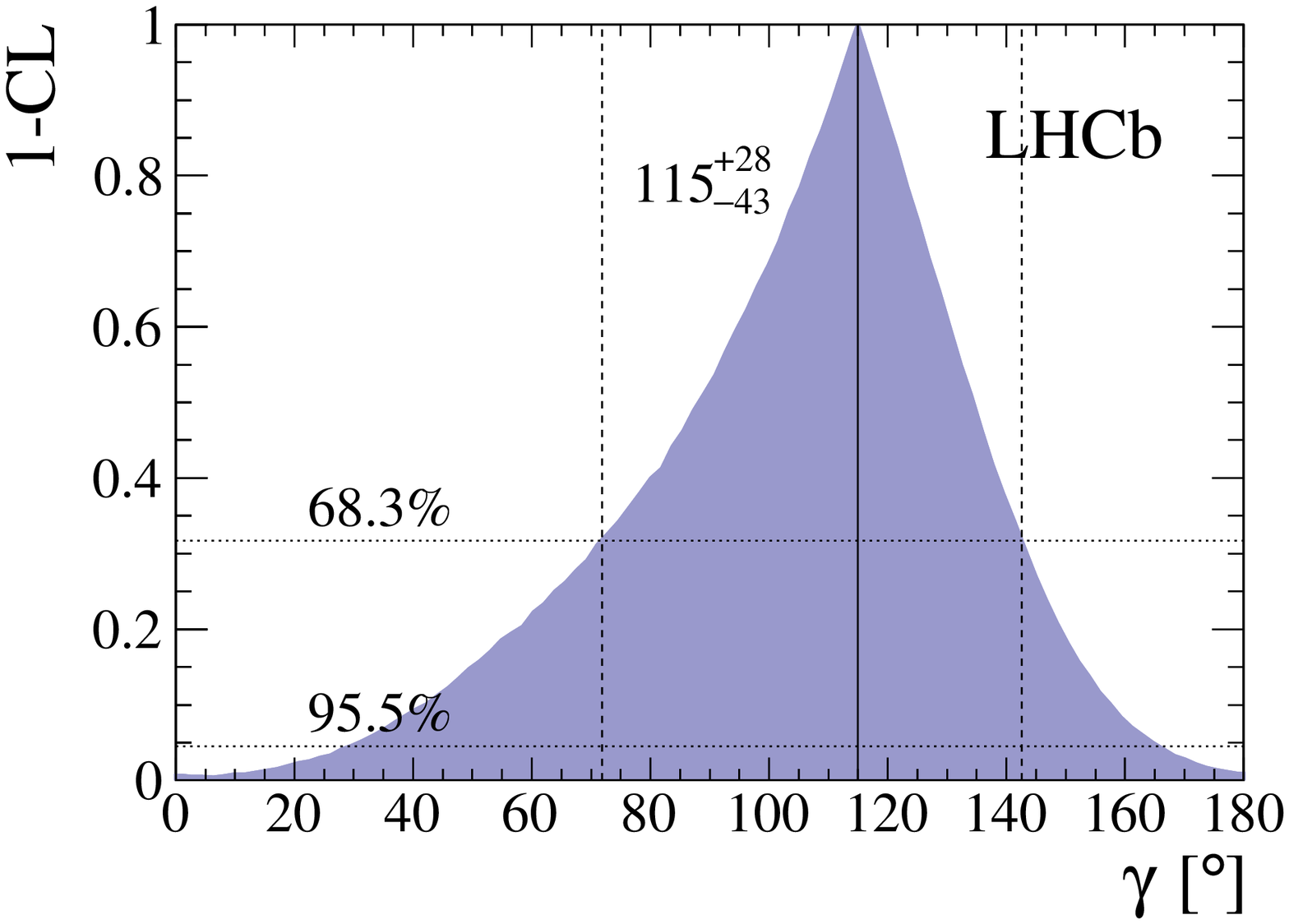}
 \includegraphics[width=.30\textwidth]{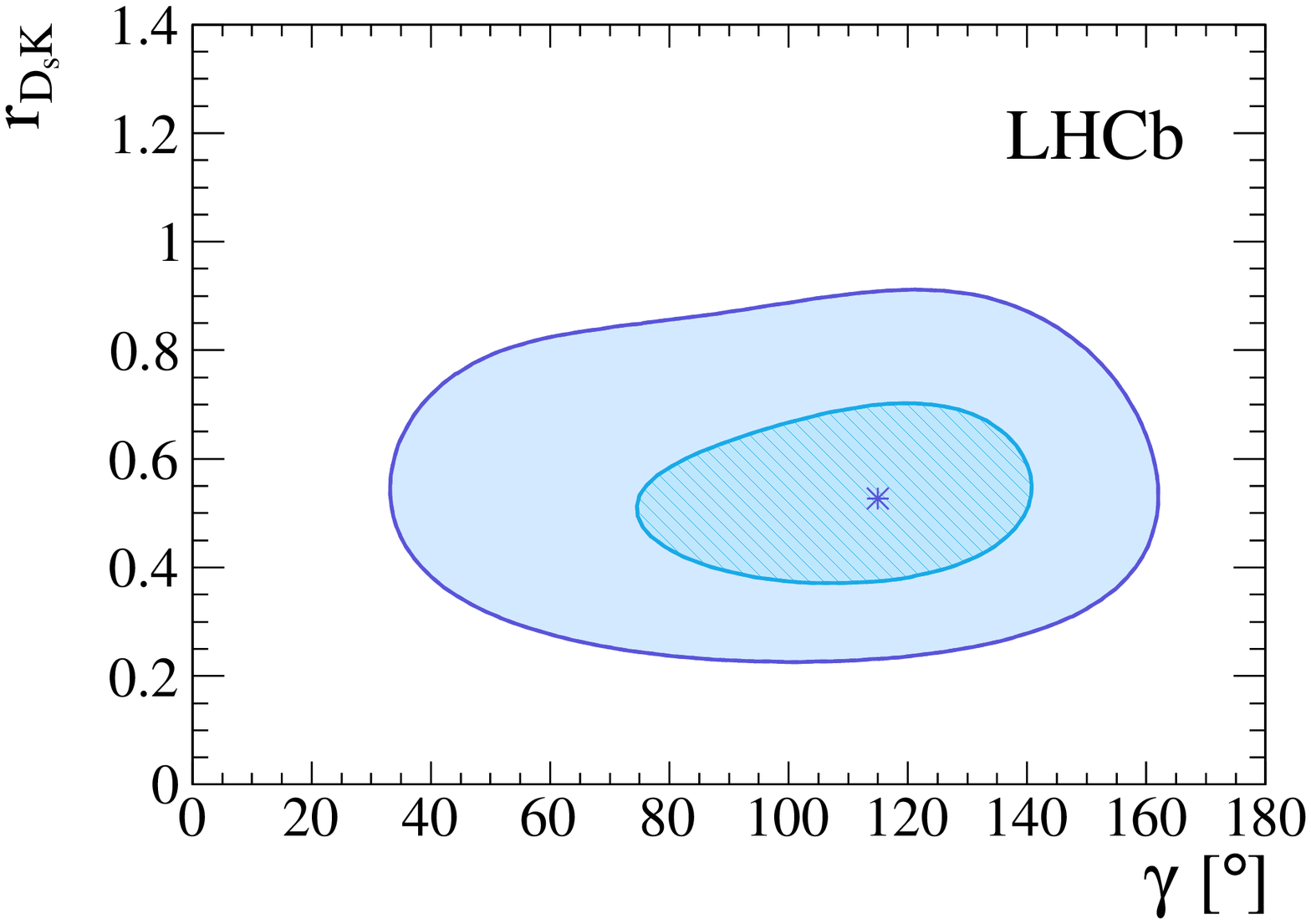}
 \includegraphics[width=.30\textwidth]{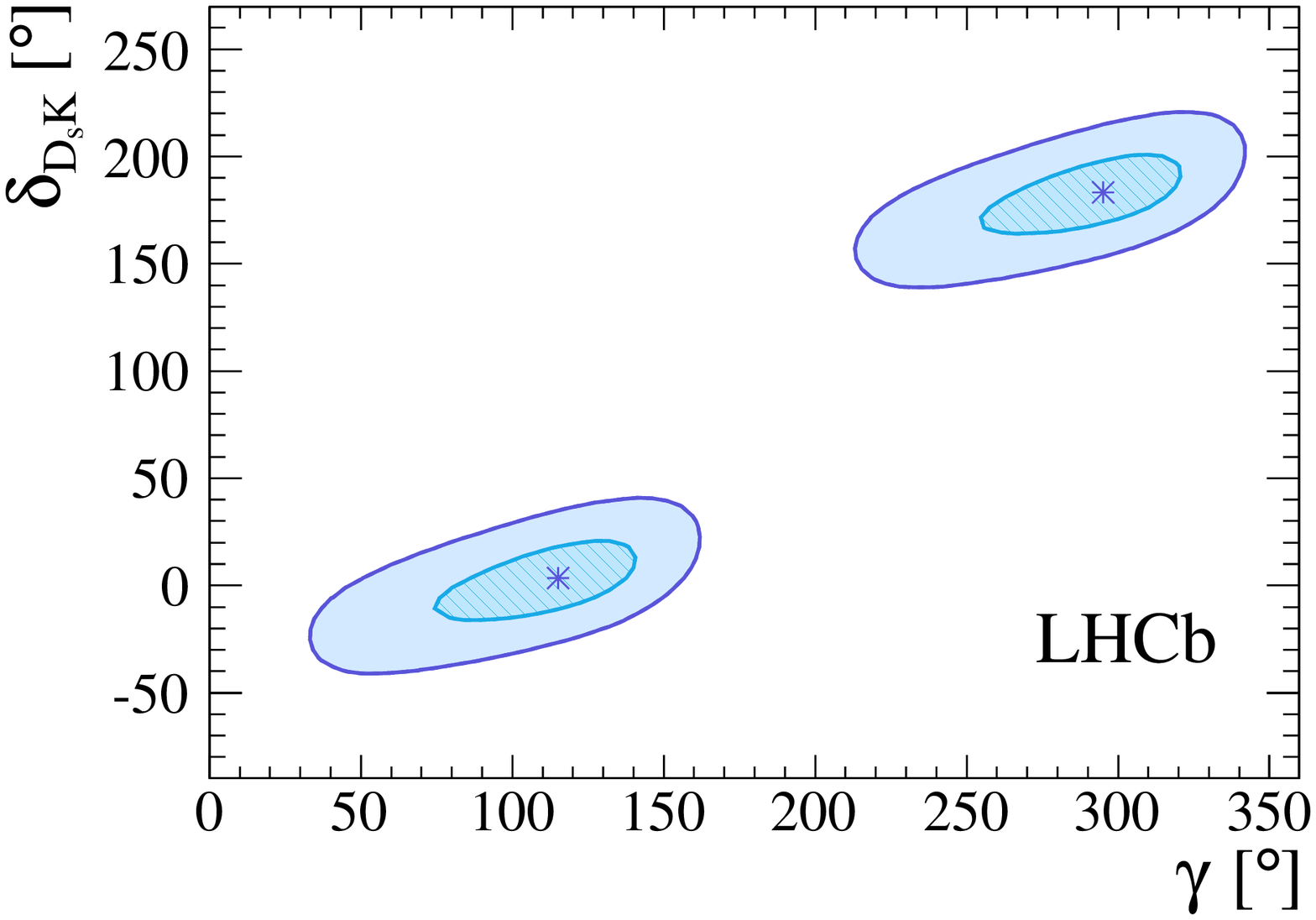}
 \caption{1-CL for $\gamma$, together with the central value and the $68.3\%$ CL interval
(left). Profile likelihood contours
of $r_{D_sK}$ vs. $\gamma$ (middle), and $\delta$ vs. $\gamma$ (bottom). The contours are at $1\sigma$ ($2\sigma$),
corresponding to $39\%$ CL ($86\%$ CL) in the Gaussian
approximation. The markers denote the best-fit values.
}
 \label{fig:gamma}
\end{figure}

Confidence intervals are computed using a frequentist method, and found to be:
\begin{center}
$\gamma = (115^{+28}_{-43})^\circ$, \\
$\delta_{D_sK} = (3^{+19}_{-20})^\circ$, \\
$r_{D_sK} = 0.53^{+0.17}_{-0.16}$,
\end{center}
where the intervals for the angles are expressed modulo $180^\circ$.
Fig.~\ref{fig:gamma} shows the $1-\textrm{CL}$ curve for
$\gamma$, and the two-dimensional
contours of the profile likelihood.

\section{Conclusion}
The time-dependent $CP$ observables accessible through $B_s^0 \rightarrow D_s^{\mp}K^{\pm}$ decays have been measured for the first time
and found to be: 
$C_f = 0.53 \pm 0.25 \pm 0.04$,
$A^{\Delta\Gamma}_{f} = 0.37 \pm 0.42 \pm 0.20$, 
$A^{\Delta\Gamma}_{\overline{f}} = 0.20 \pm 0.41 \pm 0.20$, 
$S_f = −1.09 \pm 0.33 \pm 0.08$,
$S_{\overline{f}} = −0.36 \pm 0.34 \pm 0.08$,
where the first uncertainty is statistical and the second systematic. 
Using these observables, the CKM angle $\gamma$ is determined to be $(115_{-43}^{+28})^\circ$ modulo $180^\circ$ at $68\%$ CL,
where the uncertainty contains both statistical and systematic components.

\Acknowledgements
This analysis took several years and was a huge collaborative effort, and I would like to express my thanks and appreciation to the
$B_s^0 \rightarrow D_s^{\mp}K^{\pm}$ analysis team: Suvayu Ali, Agnieszka Dziurda, Till Moritz Karbach,
Rose Koopman, Eduardo Rodrigues, Manuel Schiller, Maximillian Schlupp, Giulia Tellarini, and Stefania Vecchi. I would also
like to give a special thanks to the subdetector and computing teams of the LHCb collaboration, without whom none of our physics
analyses would see the light of day.


\begin{thebibliography}{99}

\bibitem{CKM1}N. Cabibbo, %Unitary symmetry and leptonic decays, 
Phys. Rev. Lett. 10 (1963) 531.
\bibitem{CKM2}M. Kobayashi and T. Maskawa, %CP violation in the renormalizable theory of weak interaction, 
Prog. Theor. Phys. 49 (1973) 652.
%\cite{Huet:1994jb}
\bibitem{Huet:1994jb}
  P.~Huet and E.~Sather,
  %``Electroweak baryogenesis and standard model CP violation,''
  Phys.\ Rev.\ D {\bf 51} (1995) 379.
  %[hep-ph/9404302].
%\cite{Brod:2013sga}
\bibitem{Brod:2013sga}
  J.~Brod and J.~Zupan,
  %``The ultimate theoretical error on $\gamma$ from $B \to DK$ decays,''
  JHEP {\bf 1401} (2014) 051.
  %[arXiv:1308.5663 [hep-ph]].
  %%CITATION = ARXIV:1308.5663;%%
  %7 citations counted in INSPIRE as of 12 Nov 2014
%\bibitem{TheorUncert}J.Zupan, The case for measuring gamma precisely, UCHEP-11-10.
  %%CITATION = HEP-PH/9404302;%%
  %261 citations counted in INSPIRE as of 12 Nov 2014
%\bibitem{Dunietz}I. Dunietz and R. G. Sachs, Asymmetry between inclusive charmed and anticharmed modes in $B^0$ , $\overline{B}^0$ decay as a measure of CP violation, Phys. Rev. D37 (1988) 3186.
%\bibitem{Aleksan}R. Aleksan, I. Dunietz, and B. Kayser, Determining the CP violating phase $\gamma$, Z. Phys. C54 (1992) 653.
\bibitem{Fleisher}R. Fleischer, %New strategies to obtain insights into CP violation through $B_{(s)} \rightarrow D_{(s)}^{\pm}K^{\mp}, D_{(s)}^{*\pm}K^{\mp}$, ... $B_{(d)} \rightarrow D^\pm \pi^\mp, D^{*\pm} \pi^\mp$, ... decays, 
Nucl. Phys. B671 (2003) 459.
\bibitem{gligckm2010} V.~V.~Gligorov, %Time-dependent measurements of the CKM angle $\gamma$ at LHCb, 
Proceedings of CKM 2010.
\bibitem{Akiba:2008zza} K.~Akiba {\it et al.},
%``Determination of the CKM-angle gamma with tree-level processes at LHCb,''
CERN-LHCB-2008-031.
\bibitem{detector}LHCb collaboration, A. A. Alves Jr. et al., %The LHCb detector at the LHC, 
JINST {\bf 3} (2008) S08005.
\bibitem{lhcbtrigger} R. Aaij et al., %The LHCb Trigger and its Performance in 2011, 
JINST {\bf 8} (2013) P04022.
%\bibitem{BDTG} L. Breiman, J. H. Friedman, R. A. Olshen, and C. J. Stone, Classification and regression trees, Wadsworth international group, Belmont, California, USA, 1984.
%\bibitem{TMVA} A. Hocker et al., TMVA - toolkit for multivariate data analysis, PoS ACAT (2007) 040.
%\cite{Gligorov:2012qt}
\bibitem{Gligorov:2012qt} V.~V.~Gligorov and M.~Williams, %Efficient, reliable and fast high-level triggering using a bonsai boosted decision tree, 
JINST {\bf 8} (2013) P02013
\bibitem{LHCbTagging}M.Dorigo, %$b$-flavour tagging in $pp$ collisions, 
Proceedings of ICHEP 2014. 
%\cite{Karbach:2014qba}
\bibitem{Manuel}
  T.~M.~Karbach, G.~Raven and M.~Schiller,
  %``Decay time integrals in neutral meson mixing and their efficient evaluation,''
  arXiv:1407.0748.
  %%CITATION = ARXIV:1407.0748;%%
  %1 citations counted in INSPIRE as of 12 Nov 2014
\bibitem{deltaMS}LHCb collaboration, R. Aaij et al., %Precision measurement of the $B_s^0 - \overline{B}_s^0$ oscillation frequency in the decay $B_s^0 \rightarrow D_s^{-}\pi^{+}$ , 
New J. Phys. 15 (2013) 053021,
\bibitem{PDG}Particle Data Group, J. Beringer et al., %Review of particle physics, 
Phys. Rev. D86 (2012) 010001.
\bibitem{Gamma}LHCb collaboration, R. Aaij et al., %Measurement of CP violation and the $B_s^0$ meson decay width difference with $B_s^0 \to J/\phi K^+K^-$ and $B_s^0 \to J/\phi \pi^+\pi^-$ decays, 
Phys. Rev. D87 (2013) 112010.
\bibitem{sPlot}M. Pivk and F. R. Le Diberder, %sPlot: a statistical tool to unfold data distributions, 
Nucl. Instrum. Meth. A555 (2005) 356.
%\bibitem{LHCbGamma}LHCb collaboration, R. Aaij et al., %A measurement of the CKM angle $\gamma$ from a combination of $B^\pm \rightarrow Dh^\pm$ analyses, 
%Phys. Lett. B726 (2013)

\end{thebibliography}
\end{document}